# Lattice-dynamics effects on small-polaron properties

Marco Zoli*

*Dipartimento di Matematica e Fisica, Istituto Nazionale di Fisica della Materia, Universitá di Camerino, 62032 Camerino, Italy*

This study details the conditions under which strong-coupling perturbation theory can be applied to the molecular crystal model, a fundamental theoretical tool for analysis of the polaron properties. I show that lattice dimensionality and intermolecular forces play a key role in imposing constraints on the applicability of the perturbative approach. The polaron effective mass has been computed in different regimes ranging from the fully antiadiabatic to the fully adiabatic. The polaron masses become essentially dimension independent for sufficiently strong intermolecular coupling strengths and converge to much lower values than those traditionally obtained in small-polaron theory. I find evidence for a self-trapping transition in a moderately adiabatic regime at an electron-phonon coupling value of $\simeq 3$. Our results point to a substantial independence of the self-trapping event on dimensionality.

## I. INTRODUCTION

The study of polaron properties has become a significant branch in condensed matter physics after Landau introduced the concept of an electron which can be trapped *by digging its own hole* in an ionic crystal.[1] Since then, several investigations[2–13] have analyzed the conditions under which polarons can form, their extension in real and momentum space, and the features of their motion both in physical[14] and biological[15,16] systems. While in general a sizable *electron-phonon coupling* is a requisite for polaron formation, also the *dimensionality* and *degree of adiabaticity* of the physical system could essentially determine the stability and behavior of the polaronic quasiparticle. When the characteristic phonon energy $\hbar \bar{\omega}$ is smaller than the electronic energy $D$ the system is set in the adiabatic regime. In these conditions traditional polaron theory[7,17,18] finds in one dimension (1D) a continuous transition between a large polaron at weak *e*-ph coupling and a small polaron at strong *e*-ph coupling, the polaron solution being in any case the ground state of the system. This picture changes drastically in 2D and 3D since a minimum value of the *e*-ph coupling is required to form finite-size polarons. The crossover between the infinite- and finite-radius polarons locates the *self-trapping* transition which should be signaled by an abrupt increase in the 2D (and 3D) quasiparticle effective mass. Being understood that the ground-state energy is an analytic function of the *e*-ph coupling parameter[19] and therefore the self-trapping process is not a phase transition, it is clear that the physical properties of the system crucially depend on the width and weight of the polaron. In view of the importance of these issues and also in connection with the present debate on the possibility of polaronic mechanisms in high-$T_c$ superconductors,[20–22] we have undertaken a reexamination of the perturbative approach to the Holstein molecular crystal model[3] in the *intermediate-to strong*-coupling regime. The Holstein model (and the extensions of the model incorporating the spin degrees of freedom) has experienced a surge of interest in the last years, and different methods, both analytical and numerical, have revealed the richness and complexity of the polaron physics both in the ground state and at finite temperatures.[23–32] In particular a recent study[33] has pointed out that a realistic on-site potential with anharmonic features can change considerably the size of the lattice deformation carried by the electron in two dimensions. These findings emphasize the relevance of the lattice for reliable predictions on the polaron motion which can be extracted from the Holstein model. In some previous works I have shown how the range and the strength of the *interatomic (intermolecular) forces* strongly affect fundamental polaron properties such as the bandwidth[34] and the effective mass in an antiadiabatic regime. This paper extends the perturbative analysis of the Holstein model to the more delicate and controversial adiabatic regime, pointing out the limits of applicability of the method and their dependence on the system dimensionality. The traditional distinction between 1D behavior on the one hand and 2D (3D) behavior on the other hand is questioned by our investigation of the adiabatic polaron, thus supporting the existence of a self-trapping transition also in 1D. In Sec. II, we review the main results of strong-coupling perturbative theory and the numerical results are displayed in Sec. III. Section IV contains some concluding remarks.

## II. HOLSTEIN MODEL HAMILTONIAN

The Hamiltonian for the single electron in the Holstein model reads

$$H = -t \sum_{i \neq j} c_i^\dagger c_j + g \sum_i c_i^\dagger c_i (a_i + a_i^\dagger) + \sum_{\mathbf{k}} \omega_{\mathbf{k}} a_{\mathbf{k}}^\dagger a_{\mathbf{k}}, \quad (1)$$

where the dimension dependence explicitly appears in the momentum space Hamiltonian for the harmonic lattice vibrations. $c_i^\dagger$ ($c_i$) creates (destroys) a tight-binding electron at the $i$ site and $t$ is the first-neighbor hopping integral related to the bare electron half bandwidth $D$ by $D = zt$, $z$ being the coordination number. $a_{\mathbf{k}}^\dagger$ ($a_{\mathbf{k}}$) creates (destroys) a $\mathbf{k}$ phonon with frequency $\omega_{\mathbf{k}}$. $g$ is the electron-phonon coupling constant. The Holstein model was originally proposed[3] in the form of a discrete nonlinear Schrödinger equation (DNLSE) for electrons whose probability amplitudes at the molecular lattice sites depend on the internuclear coordinates of the diatomic molecules. The nonlinearity parameter which accounts for the energy gain due to polaron formation ($A$ in Ref. 3 but often denoted by $\chi$ in the literature) is linked to $g$ by the relation $g^2 = dA^2\hbar/2M\omega_0$, $d$ being the system dimension, $M$ the reduced molecular mass, and $\omega_0$ the intramolecular breathing mode frequency. Since $g$ scales $\propto \sqrt{d}$, it is convenient to introduce the $d$-independent $e$-ph coupling $g_0 \equiv g/\sqrt{d}$ which will be used throughout this paper in units of $\hbar\omega_0$. The ratio between the polaron binding energy $E_b^0$ and the electron kinetic energy $D$ is also dimension independent, being $\lambda^0 \equiv E_b^0/D = dA^2/2M\omega_0^2 zt$, with $d/z = 1/2$ both for the linear chain, the square lattice, and the simple cubic lattice. The apex 0 refers to a model with dispersionless phonons.

Here we only note that the connection between the Holstein Hamiltonian and DNLSE is not trivial,[35] the DNLSE being obtainable from the Holstein Hamiltonian under (i) the semiclassical approximation which decouples the fluctuations of the fermionic and bosonic fields and (ii) the adiabatic approximation which assumes an atomic (bosonic) motion enslaved to that of the electrons (fermions). In the antiadiabatic limit $\hbar\bar{\omega} \gg D$ the use of the DNLSE may be therefore questionable.

The strong-coupling perturbative method can be applied when the condition $D < g$ is fulfilled.[6,36] This already implies that the method works better in 1D and worse in 3D. The Lang-Firsov unitary transformation[5] assumes the strong-coupling regime and transforms $H$ of Eq. (1) into $\tilde{H} = \tilde{H}_0 + \tilde{H}_P$ with

$$\tilde{H}_0 = \sum_{\mathbf{k}} \hbar\omega_{\mathbf{k}} a^\dagger_{\mathbf{k}} a_{\mathbf{k}} - \frac{g^2}{N} \sum_{\mathbf{k}} (\hbar\omega_{\mathbf{k}})^{-1} \sum_j c^\dagger_j c_j$$
$$- \frac{g^2}{N} \sum_{\mathbf{k}} \sum_{i \neq j} \frac{\exp[i\mathbf{k}\cdot(\mathbf{r}_i - \mathbf{r}_j)]}{\hbar\omega_{\mathbf{k}}} c^\dagger_j c_j c^\dagger_i c_i,$$

$$\tilde{H}_P = -t \sum_{i \neq j} \exp\left[-\frac{2g^2}{N}\sum_{\mathbf{k}}(\hbar\omega_{\mathbf{k}})^{-2}\sin^2\frac{[\mathbf{k}\cdot(\mathbf{r}_i - \mathbf{r}_j)]}{2}\right]$$
$$\times \sum_{m=0}^\infty \frac{1}{m!}\left[\frac{g}{\sqrt{N}}\sum_{\mathbf{k}} \frac{a^\dagger_{-\mathbf{k}}}{\hbar\omega_{\mathbf{k}}}(e^{i\mathbf{k}\cdot\mathbf{r}_i} - e^{i\mathbf{k}\cdot\mathbf{r}_j})\right]^m$$
$$\times \sum_{n=0}^\infty \frac{1}{n!}\left[\frac{g}{\sqrt{N}}\sum_{\mathbf{k}} \frac{a_{\mathbf{k}}}{\hbar\omega_{\mathbf{k}}}(e^{i\mathbf{k}\cdot\mathbf{r}_j} - e^{i\mathbf{k}\cdot\mathbf{r}_i})\right]^n c^\dagger_i c_j, \quad (2)$$

where $\mathbf{r}_i$ and $\mathbf{r}_j$ are the lattice vectors of neighboring molecular sites. $N$ is the number of diatomic molecules in the lattice. The starting problem of a strongly $e$-ph interacting system is thus mapped onto that of *dressed electrons* interacting with phonons via $\tilde{H}_P$. Here $\tilde{H}_0$ is diagonal except for a small term corresponding to an attractive polaron-polaron interaction.[37] The usual perturbative procedure calculates the contribution to the ground-state polaron band $E(\mathbf{p})$, $\mathbf{p}$ being the total crystal momentum, by applying $\tilde{H}_P$ to the eigenstates of $\tilde{H}_0$, averaging over the phonon variables, and keeping the vacuum state without phonons.[38] Then, to first order, only the $m = n = 0$ term in $\tilde{H}_P$ yields a nonzero contribution to $E(\mathbf{p})$ and the polaron hopping from the $j$th to the $i$th site is not affected by self-energy corrections due to phonon exchange. Instead, to second order, when the polaron hops from $j$ to $i$, intermediate states having $1, \ldots, m$ $\mathbf{k}$ phonons are created via application of $\tilde{H}_P$ to the vacuum. Summation over such intermediate states implies therefore that the second-order polaron self-energy comprises the emission and absorption of an arbitrary number of phonons. Moreover, the second hop leads the polaron from the intermediate $i$ site to the final $f$ site which either coincides with the starting $j$ or is two lattice spacings distant from $j$. While the former process simply renormalizes the polaron binding energy, the latter does introduce dispersive features in the ground-state band which turn out to be dimension dependent since higher-dimension systems are more closely packed.

### III. NUMERICAL RESULTS

The polaron mass $m^*$ is defined by

$$\frac{m^*}{m_0} = \frac{zt}{\nabla^2 E(\mathbf{p})|_{\mathbf{p}=0}}, \quad (3)$$

where $m_0$ is the bare band mass. I have calculated $E(\mathbf{p})$ according to the second-order perturbative procedure outlined in the previous section and deduced the following general expression for the effective mass:

$$\left(\frac{m^*}{m_0}\right)_d = \frac{\exp(\bar{g}^2)}{1 + z^2 t \exp(-\bar{g}^2) f(\bar{g}^2)/(\hbar\bar{\omega})},$$

$$\bar{g}^2 = \frac{2g^2}{N} \sum_{k_x} \sin^2\frac{k_x}{2} \sum_{k_x,k_y} (\hbar\omega_{\mathbf{k}})^{-2},$$

$$f(\bar{g}^2) = \sum_{s=1}^{+\infty} \frac{(\bar{g}^2)^s}{ss!}. \quad (4)$$

The series expansion in the last equation of Eqs. (4) reflects the fact that the second-order polaron self-energy is a sum over an infinite number of diagrams, each having $s$ phonons between the two interaction vertexes. It can be computed using the following relation:[39]

$$f(\bar{g}^2) = \mathrm{Ei}(\bar{g}^2) - \gamma - \ln(\bar{g}^2) = \int_0^{\bar{g}^2} dt \frac{\exp(t) - 1}{t}, \quad (5)$$

where $\mathrm{Ei}(x)$ is the exponential integral and $\gamma$ is the Euler constant. Note that the second order of perturbative theory also introduces the effect of the adiabaticity parameter on $m^*$. Moreover, $m^*$ depends on dimensionality through (i) $g^2 \propto d$, (ii) the first-neighbor number $z$, (iii) the Brillouin zone sums, and (iv) the features of the phonon spectrum. We have assumed pairwise intermolecular interactions and deduced the analytic phonon frequencies for the linear chain (1D), the square lattice (2D), and the simple cubic lattice (3D).[34] Here I restrict the range of the intermolecular forces to the first-neighbor molecular sites so that two parameters control the phonon spectrum.[40] the intramolecular force constant $\beta$ and the intermolecular first-neighbor force constant $\gamma$ in terms of which one defines $\omega_0^2 = 2\beta/M$ and $\omega_1^2 = \gamma/M$. Then the zone center frequency is $\bar{\omega} = \sqrt{\omega_0^2 + z\omega_1^2}$.

My previous works have pointed out that the Holstein model with a dispersionless spectrum ($\omega_1=0$) or with weak intermolecular forces ($\omega_1 \ll \omega_0$) is erroneous since (a) it would predict *larger polaron bandwidths in lower dimensionality* and (b) it would yield a *divergent site jump probability* for the polaronic quasiparticle. Numerical analysis has shown that the bandwidths $\Delta E_d$ grow faster versus $\omega_1$ in higher dimensionality, thus providing a criterion to fix the minimum $\omega_1$ which ensures the validity of the Holstein model. The *threshold value* $\bar{\omega}_1$ has been in fact determined according to the inequalities criterion $\Delta E_{3D} \geq \Delta E_{2D} \geq \Delta E_{1D}$ and $\bar{\omega}_1$ has proved to be essentially dependent on the breathing mode frequency $\omega_0$ and on the *e*-ph coupling $g_0$. Summing up, at intermediate $g_0$ ($\simeq 1-1.5$), $\bar{\omega}_1$ is $\simeq \omega_0/2$, while at strong $g_0$ ($\geq 2$), $\bar{\omega}_1$ should be at least $\simeq 2\omega_0/3$ in order to guarantee the correct bandwidth trend. It follows that also the polaron mass values obtained in the Holstein model are reliable *only if* the intermolecular coupling strengths are sufficiently strong. For these reasons the $m^*$-$\omega_1$ plots hereafter displayed do not contain the lower portion of $\omega_1$ values.

Although the present work is mainly concerned with the constraints on the applicability of the perturbative method to the Holstein model, the figures we put in the numerical analysis are appropriate to some strong-coupling systems in which polaronic features have been envisaged. Consistently with some previous studies[30] we compute the masses, taking $g_0=1.3$ as the minimum coupling value which ensures the breakdown of the Migdal theorem.

Let us start the discussion from an antiadiabatic regime characterized by $t=15$ meV and $\omega_0=100$ meV. In Fig. 1, the ratios $(m^*/m_0)_d$ are plotted versus $\omega_1$ in 1D, 2D, and 3D by assuming two *e*-ph couplings $g_0$: an intermediate value $g_0=1.3$ (see the lower curves in each figure) and a strong value $g_0=2.2$ (upper curves in each figure). The former case presents $\bar{\omega}_1=48$ meV while in the latter $\bar{\omega}_1=62$ meV. In both cases the first-order and the second-order (first-plus second-order term) perturbative results are reported on. The second-order correction lowers the mass values in any dimensionality with particular significance in 3D. Moreover, the second-order correction is a growing function of $\omega_1$ and its effectiveness is larger in the intermediate *e*-ph coupling case. When the second- and first-order terms become comparable the perturbative method breaks down: in 3D with $g_0=1.3$ this event takes place at $\omega_1=58$ meV as the arrow signals in the lower part of Fig. 1. Considering that the threshold value in the intermediate-coupling case is $\bar{\omega}_1=48$ meV, we observe that the perturbative approach to the Holstein model has a restricted $\omega_1$ range of validity given by $48\ \text{meV} \leq \omega_1 \leq 58\ \text{meV}$. Here $\omega_1 < 48$ meV implies the failure of the Holstein model in itself while at $\omega_1 > 58$ meV it is the perturbative method that fails. Of course, by enhancing $g_0$ the perturbative method works better as the absence of arrows in the upper curves of Fig. 1 points out.

We emphasize that the breakdown of the perturbative method is closely related to the inadequacy of the Lang-Firsov approach:[41,42] when the electronic and phononic subsystems are not strongly coupled the lattice deformation does not follow coherently the electron through the crystal and the polaronic unit broadens in real space. Then the crossover between a small-polaron (at strong $g_0$) and a large-polaron (at intermediate $g_0$) solution shows up in a decreased lattice deformation and associated lowering of the energy gain due to polaron formation: under these conditions the Lang-Firsov scheme becomes less appropriate. Finally note that the mass ratios converge to substantially *d*-independent values in the upper portion of the $\omega_1$ range being, for any *d*, $m^*/m_0 \simeq 3$ at $\omega_1 \simeq 60$ meV in the $g_0=1.3$ case and $m^*/m_0 \simeq 10$ at $\omega_1 \simeq 80$ meV in the $g_0=2.2$ case.

Next we take a larger hopping integral, $t=45$ meV in Fig. 2, thus increasing the degree of adiabaticity. This choice implies in fact that the 1D system is still antiadiabatic ($\hbar\bar{\omega}/2t > 1$), the 2D system is in *antiadiabatic to adiabatic* transition regime according to the strength of $\omega_1$, and the 3D system is already in adiabatic conditions. A larger $t$ also requires a stronger *e*-ph coupling in order to fulfill the perturbative condition. We see that the perturbative method works well in 1D and 2D while, in 3D with $g_0=2.2$, it fails at $\omega_1 \geq 70$ meV. Even larger *e*-ph couplings ($g_0 \simeq 3$ in the upper curves) are necessary to balance the effect of the intermolecular forces in the second-order correction and recover the correctness of the perturbative approach. The threshold values for the validity of the Holstein model are set in Fig. 2 at $\bar{\omega}_1=62$ meV (lower curves) and $\bar{\omega}_1=67$ meV (upper curves), respectively. Again, a substantial *d* independence of the mass values is achieved at large $\omega_1$: $m^*/m_0 \simeq 10$ at $\omega_1 \simeq 70$ meV in the $g_0=2.2$ case and $m^*/m_0 \simeq 100$ at $\omega_1 \simeq 80$ meV in the $g_0=3.1$ case.

Let us turn to a moderately adiabatic case with $t=\omega_0=50$ meV. Since the intermolecular forces are expected not to exceed the intramolecular forces in real systems, we set $\omega_1 \leq 50$ meV in Fig. 3. The perturbative approach applies in both 1D strong-coupling cases, while it fails in 2D with $g_0=2.6$ at $\omega_1 \geq 40$ meV. In 3D, two arrows show up at $\omega_1=33$ meV (case $g_0=2.6$) and $\omega_1=46$ meV (case $g_0=3.6$), respectively. Considering that the threshold $\bar{\omega}_1$ is exactly 33 meV in the former case it turns out that the perturbative approach has no range of validity in 3D with $g_0=2.6$ and larger $g_0$ values are necessary to sustain the method. The 1D and 2D mass ratios converge to $\simeq 10$ at $\omega_1 \simeq 40$ meV in the $g_0=2.6$ case and to $\simeq 30$ at $\omega_1 \simeq 50$ meV in the $g_0=3.6$ case. We note that the 1D $g_0=2.6$ case applies well[43] to the system tetrathiafulvalene-tetracyanoquinodimethane (TTF-TCNQ), a molecular crystal with 1D conduction properties, and a small-polaron binding energy of $\simeq 0.2$ eV. This means, in our model with $\omega_0=50$ meV, an intermolecular $\omega_1 \simeq 40$ meV which yields a 1D mass ratio of $\simeq 10$. This value is consistent with recent variational studies[23] and suggests that the current carrying elementary excitations for the TTF-TCNQ system are one order of magnitude lighter than estimated in the past.[44] The 3D $g_0=2.6$ case is appropriate instead to a classical polaronic system, $SrTiO_3$. At the largest intermolecular energy consistent with the perturbative method, we find a mass ratio of $\simeq 50$, a factor of 2 smaller than previously reported in small-polaron theory but still somewhat higher than the observed value ($\simeq 20$).[45] This would confirm[6] that rather an intermediate-polaron description is suitable to this perovskite structure.[12]

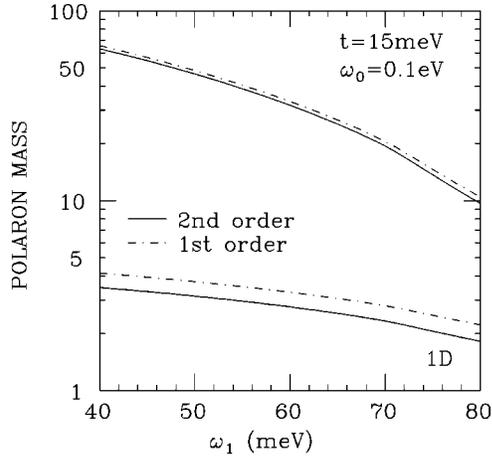
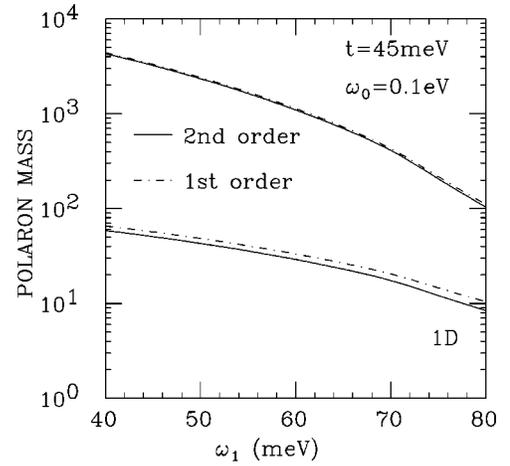
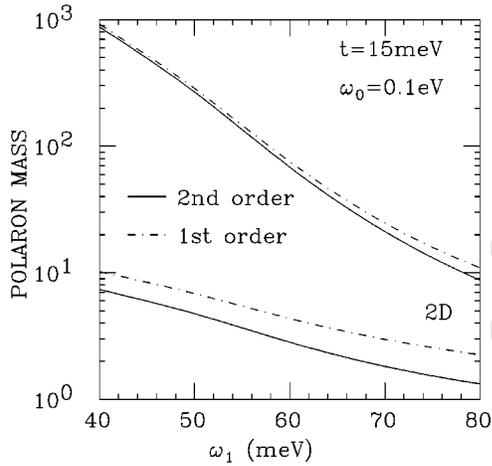
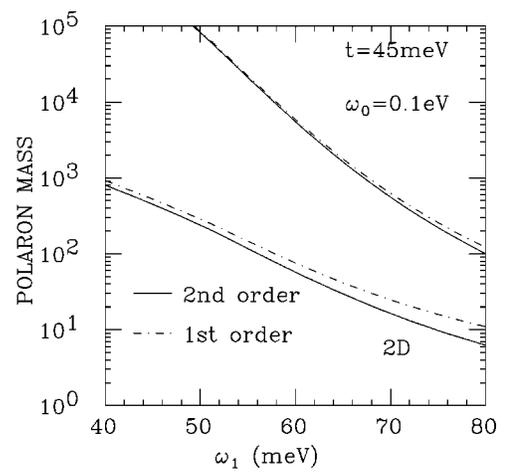
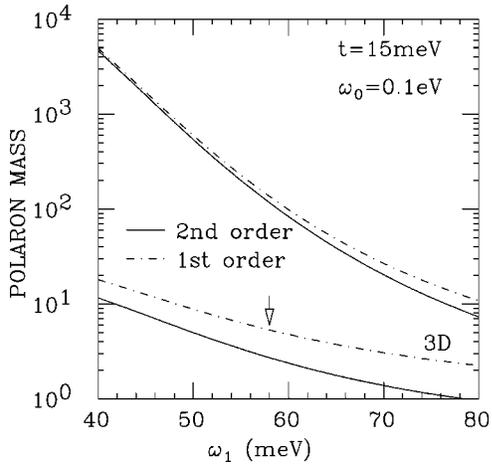
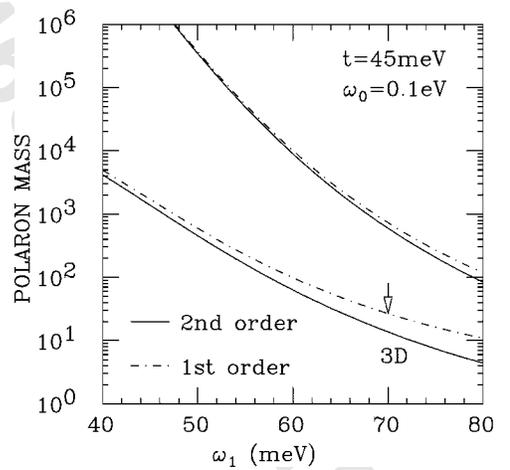

FIG. 1. Polaron masses (in units of the bare band mass) vs the first-neighbor intermolecular energy in 1D, 2D, and 3D. The hopping integral is $t=15$ meV and the intramolecular frequency is $\omega_0=100$ meV. Both the first- and second-order perturbative results are displayed for two choices of electron-phonon coupling constant: $g_0=1.3$ (lower curves) and $g_0=2.2$ (upper curves). The arrow in the 3D $g_0=1.3$ case marks the minimum value of intermolecular coupling at which first- and second-order perturbative terms become comparable; hence, the perturbative approach fails at larger $\omega_1$. The absence of arrows in the other cases indicates that the perturbative method works throughout the whole range of intermolecular couplings.

FIG. 2. As in Fig. 1 but with $t=45$ meV. The lower curves in each figure refer to an electron-phonon coupling $g_0=2.2$ while the upper curves have been obtained with $g_0=3.1$. In the 3D $g_0=2.2$ case, the minimum value of intermolecular coupling at which first- and second-order perturbative terms become comparable is $\omega_1=70$ meV.

Once the system is driven to a fully adiabatic regime ($t=2\omega_0$) the number of arrows grows (see Fig. 4) pointing to a substantial inadequacy of the perturbative approach unless very strong e-ph coupling conditions ($g_0 \geq 4$) had to hold. We see, for instance, that the perturbative approach cannot be used at all in the 2D $g_0=2.6$ case while in the 2D $g_0$

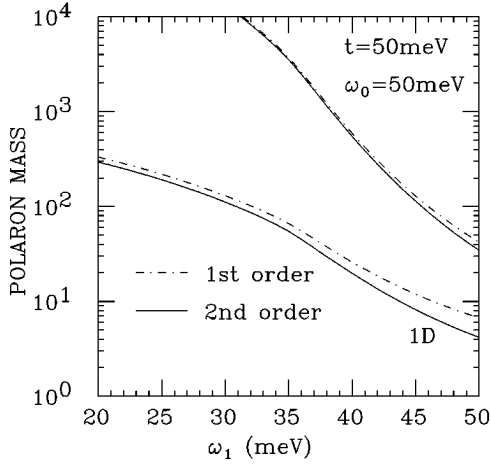
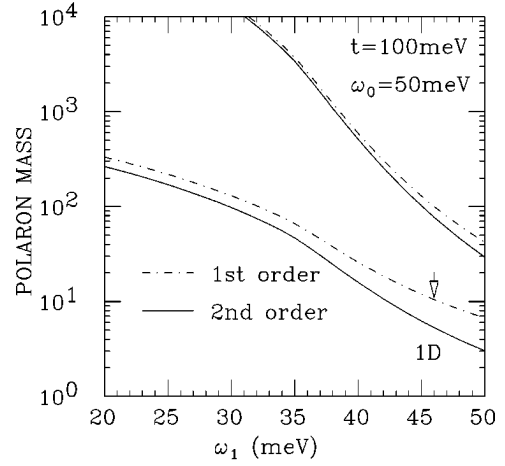
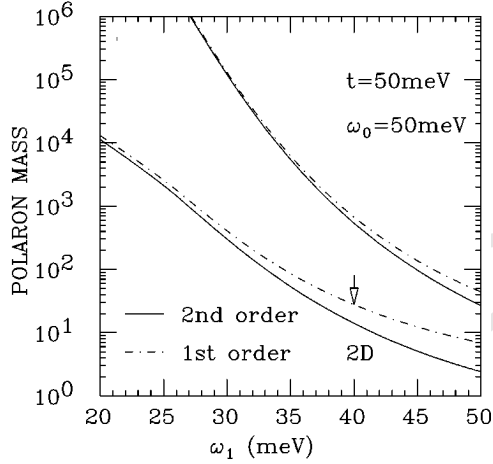
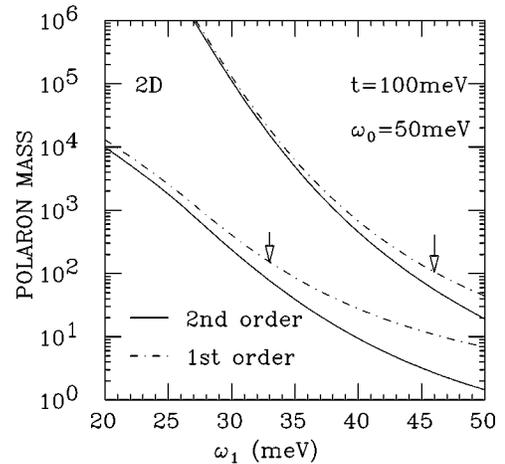
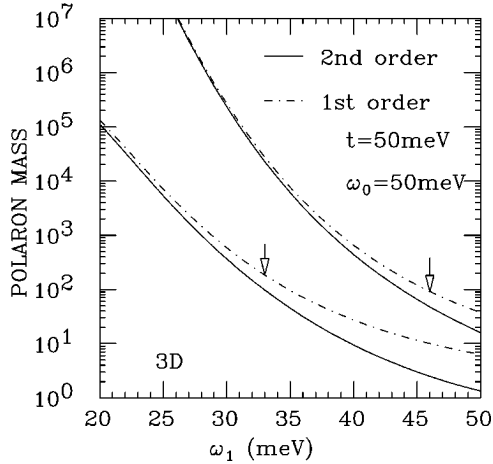
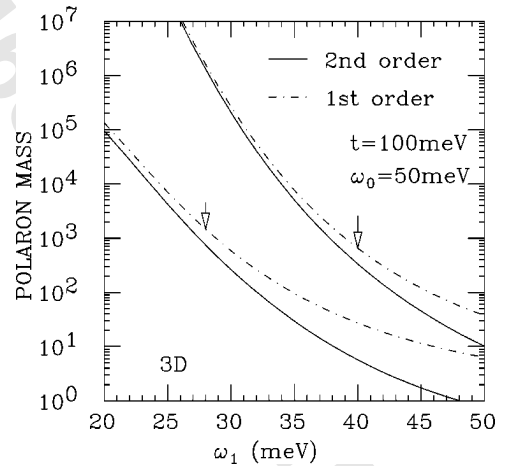

FIG. 3. Polaron masses vs $\omega_1$ in an intermediate adiabatic regime with $t=\omega_0=50$ meV. The lower curves in each figure refer to an electron-phonon coupling constant $g_0=2.6$ while the upper curves refer to $g_0=3.6$. The perturbative method is valid (i) for any $\omega_1$ in 1D, (ii) below $\omega_1=40$ meV in 2D with $g_0=2.6$, (iii) for any $\omega_1$ in 2D with $g_0=3.6$, (iv) below $\omega_1=33$ meV in 3D with $g_0=2.6$, and (v) below $\omega_1=46$ meV in 3D with $g_0=3.6$.

FIG. 4. Polaron masses vs $\omega_1$ in an adiabatic regime with $t=2\omega_0=100$ meV. The electron-phonon coupling constants are as in Fig. 3. The perturbative method is valid (i) below $\omega_1=46$ meV in 1D with $g_0=2.6$, (ii) for any $\omega_1$ in 1D with $g_0=3.6$, (iii) below $\omega_1=33$ meV in 2D with $g_0=2.6$, (iv) below $\omega_1=46$ meV in 2D with $g_0=3.6$, (v) below $\omega_1=28$ meV in 3D with $g_0=2.6$, and (vi) below $\omega_1=40$ meV in 3D with $g_0=3.6$.

$=3.6$ case, being $\bar{\omega}_1=35$ meV, the condition $35$ meV$\leq\omega_1$ $\leq 46$ meV needs to be fulfilled. In the 3D $g_0=3.6$ case, the range of applicability of the perturbative method is rather poor: $35$ meV$\leq\omega_1\leq 40$ meV. Only in the 1D $g_0=3.6$ case does the perturbative method work properly and lead to a mass ratio of $\simeq 30$ at $\omega_1\simeq 50$ meV. Note that the breakdown of the Lang-Firsov method in the adiabatic regime is signaled by a reduction of $\lambda=E_b/zt$ to values smaller than 1: when this happens the conditions for the validity of the Migdal approximation are restored.[46]

The figures for the adiabatic parameter given in Fig. 4 are appropriate to the high-$T_c$ cuprates[47] which also exhibit strong coupling of the charge carriers to some selected vibrational modes[21,48] with associated small-polaron formation[49] in the Cu-O planes. Our calculations predict a mass of $\simeq 80 m_e$ for the 2D small polaron (at the largest $\omega_1$ values consistent with the perturbative method) while lower masses are attainable once the polaron broadens in space. However, considering that slight variations of the breathing mode and intermolecular frequencies lead to relevant changes in the estimated masses, we feel that more refined computations accounting for the lattice structure of the cuprates are required, this in view of the relevance of precise charge carrier mass estimates for the theoretical discussion on polaronic high-$T_c$ superconductivity. To make a comparison with a recent path integral quantum Monte Carlo analysis[13] of the dispersionless Holstein Hamiltonian we note that, for the adiabatic case $t = 2\omega_0$, a 1D mass ratio of $\simeq 10$ is achieved in Ref. 13 when $\lambda = 1.75$. In our model, $\lambda$ does depend both on the intermolecular forces and $g_0$. For instance, given $g_0 = 3.6$ as in Fig. 4, we find that $\lambda = 1.75$ implies $\omega_1 \simeq 45$ meV (within the range in which the perturbative method works), hence a 1D mass ratio of $\simeq 100$. Moreover, unlike Ref. 13, my 2D small Holstein polaron is not heavier (rather slightly lighter) than the 1D one and we attribute this discrepancy to the absence of a dispersive phonon spectrum in the quoted Monte Carlo calculations. Finally, as for the dependence of $m^*$ on the adiabatic parameter $zt/\bar{\omega}$ [see Eq. (4)] at a given $\lambda$, we remark that two descriptions may occur: (a) for a fixed $t$, by *decreasing* $\bar{\omega}$, $m^*$ gets heavier; (b) for a fixed $\bar{\omega}$, by *increasing* $t$, $m^*$ gets lighter. Hence, for a given adiabatic ratio, the $m^*$ values can differ significantly according to the absolute values of $t$ and $\bar{\omega}$, the $m^*$ dependence on the latter being much stronger.

The self-trapping transition in polaronic systems is generally associated with an abrupt crossover between an extended state at small $e$-ph coupling and a finite-radius polaron at large $e$-ph coupling. The larger $g_0$, the smaller is the number of lattice sites over which the polaron wave function extends. When the electron digs its own hole and drags the lattice deformation unavoidably it becomes heavy so that the effective mass behavior is a clear indicator of the self-trapping transition. In antiadiabatic conditions the perturbative method predicts the absence of self-trapping in 1D, $m^*$ being a smoothly increasing function of $g_0$. Let us consider here the moderately adiabatic regime treated in Fig. 3. As the intermolecular forces strongly affect $m^*$ we need a consistent criterion to point out the features of the $m_d^*$ versus $g_0$ dependence: at any $g_0 \geq 1$ we evaluate the $d$-dependent effective masses at the previously widely discussed threshold values $\bar{\omega}_1(g_0)$, being aware that this choice guarantees the validity of the perturbative approach at least in 1D and 2D. Figure 5 displays the $m^*(\bar{\omega}_1)/m_0 - g_0$ plot in 1D and 2D.

While the mass renormalization is not severe in the intermediate- to strong- ($1 < g_0 < 2.5$) coupling regime (in agreement with path integral quantum Monte Carlo simulations of the 1D Holstein model[25]), at $g_0 = 2.55$ the polaron mass starts to grow considerably over the bare band value and at $g_0 = 3.05$ an abrupt slope change shows up unambiguously both in 1D and 2D, both in the first and second orders of perturbative theory. We interpret this result as an indicator of the self-trapping transition which then, in adiabatic regimes, should not depend on the system dimensionality. Similar conclusions have been recently reached through an independent study of the adiabatic polaron based on a variational approach.[50] Note that the 2D masses turn out to be heavier than the 1D masses in Fig. 5 as the convergence of $m_d^*$ versus $d$ takes place at values somewhat larger than $\bar{\omega}_1(g_0)$. Anyway the evidence of the self-trapping event is not affected by the choice of the intermolecular coupling strengths at which the effective masses are computed provided such strengths lie within the bounds imposed by the perturbative method.

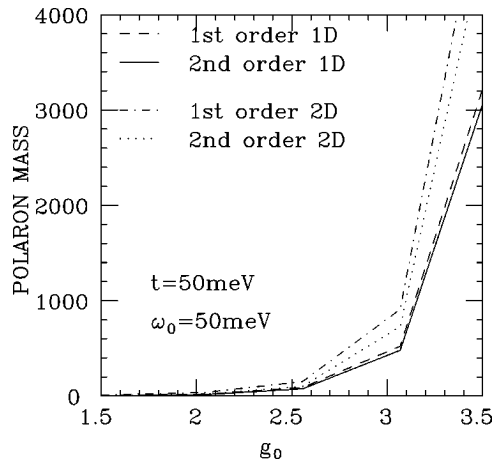

FIG. 5. 1D and 2D polaron masses vs $g_0$ in the moderately adiabatic regime with $t = \omega_0 = 50$ meV. Both the first- and second-order results of perturbative theory are displayed.

## IV. CONCLUSIONS

This study of the Holstein molecular crystal model has been motivated by the need of clarifying some open questions regarding the character of the polaronic quasiparticles in real systems, markedly the estimate of the effective mass and the existence of a self-trapping transition in low dimensionality. In spite of its apparent simplicity the Holstein Hamiltonian hosts a rich variety of polaronic behaviors according to the different physical regimes and the portions of parameters space one decides to sample. I have attacked the Holstein Hamiltonian by means of a standard perturbative method which assumes a sizable electron-phonon coupling regime in which the applicability of the Migdal approximation is ruled out. Unlike most of the previous studies on the Holstein model the lattice dimensionality and the related dispersive features of the phonon spectrum have been taken fully into account in the present investigation. The dimensionality and strengths of the intermolecular forces interfere with both the adiabatic (antiadiabatic) parameter and the $e$-ph coupling constant peculiar to the system, thus shaping the range of validity of the perturbative approach. The general rules, suggested by our numerical results, concerning the applicability of the perturbative method to the Holstein model are that (i) the method works better in strong $e$-ph coupling conditions and in an antiadiabatic regime and (ii) the method works better in low dimensionality and at *inter-* *mediate* values of the intermolecular coupling strengths. It should be immediately added that the effects of the intermolecular forces have to be treated carefully since *not sufficiently strong* intermolecular strengths would result in wrong predictions of (i) the ground-state polaronic bandwidth versus $d$, $d$ being the system dimensionality, and (ii) the polaron mobility versus temperature and $d$. Hence, the reliability of the perturbative method at weak intermolecular couplings would be a Pyrrhic victory. On the other hand, the intermolecular coupling energies have un upper bound imposed by the value of the bare intramolecular energy between the two atoms forming the basic unit in the molecular lattice model. I have computed the polaron effective mass in a number of cases, embracing a large portion of the parameter space, and presented four characteristic plots (Figs. 1–4) which essentially differ for the value of the adiabatic parameter. The computed masses are larger in Fig. 4 (adiabatic case) than in Fig. 1 (antiadiabatic case) but this result is *only apparently at variance with* the general belief according to which antiadiabatic polarons are heavier than adiabatic ones *at fixed e-ph coupling*. In fact, by enhancing $t/\omega_0$ (Fig. 1 → Fig. 4), we have also increased the *e*-ph coupling parameter to ensure the validity of the perturbative approach and this unavoidably leads to heavier masses.

When the perturbative method and the associate Lang-Firsov scheme apply correctly the polaron masses become substantially $d$ independent. This result, which has a clear character of novelty and contradicts the traditional belief of a much heavier polaron mass in low $d$, descends from the correct use of dispersive phonons in the Holstein Hamiltonian which, moreover, makes the mass renormalization less severe than predicted by dispersionless models. The mass values are, however, so sensitive to the input parameters that extreme care should be taken before applying the model predictions to real structures.

Finally, I have addressed the open question of the existence or nonexistence of a self-trapping transition in adiabatic conditions considering the nontrivial situation of a moderately adiabatic regime. The results displayed in Fig. 5 point to a positive answer corroborating the possibility of a self-trapping process also in the controversial case of one-dimensional systems.

*Electronic address: zoli@campus.unicam.it

[1] L.D. Landau, Z. Phys. **3**, 664 (1933).
[2] H. Fröhlich, Proc. R. Soc. London, Ser. A **215**, 291 (1952).
[3] T. Holstein, Ann. Phys. (N.Y.) **8**, 325 (1959); **8**, 343 (1959).
[4] Y. Toyozawa, Prog. Theor. Phys. **26**, 29 (1961).
[5] I.J. Lang and Yu.A. Firsov, Zh. Éksp. Teor. Fiz. **43**, 923 (1962) [Sov. Phys. JETP **16**, 1301 (1963)].
[6] D.M. Eagles, Phys. Rev. **145**, 645 (1966).
[7] D. Emin and T. Holstein, Phys. Rev. Lett. **36**, 323 (1976).
[8] H. De Raedt and A. Lagendijk, Phys. Rev. B **27**, 6097 (1983).
[9] H. De Raedt and A. Lagendijk, Phys. Rev. B **30**, 1671 (1984).
[10] J. Ranninger and U. Thibblin, Phys. Rev. B **45**, 7730 (1992).
[11] G. Kopidakis, C.M. Soukoulis, and E.N. Economou, Phys. Rev. B **51**, 15 038 (1995).
[12] V.N. Kostur, and P.B. Allen, Phys. Rev. B **56**, 3105 (1997).
[13] A.S. Alexandrov and P.E. Kornilovitch, Phys. Rev. Lett. **82**, 807 (1999).
[14] A.A. Gogolin, Phys. Status Solidi B **103**, 397 (1981).
[15] A.S. Davydov and N.I. Kislukha, Phys. Status Solidi B **59**, 405 (1973).
[16] D.W. Brown and Z. Ivic, Phys. Rev. B **40**, 9876 (1989).
[17] G. Kalosakas, S. Aubry, and G.P. Tsironis, Phys. Rev. B **58**, 3094 (1998).
[18] V.V. Kabanov and O.Y. Mashtakov, Phys. Rev. B **47**, 6060 (1993).
[19] H. Löwen, Phys. Rev. B **37**, 8661 (1988).
[20] For a detailed description of polaronic effects in high-$T_c$ systems, see A.S. Alexandrov and A.B. Krebs, Usp. Fiz. Nauk **163**, 1 (1992) [Sov. Phys. Usp. **35**, 345 (1992)]; A.S. Alexandrov and N.F. Mott, Rep. Prog. Phys. **57**, 1197 (1994); A.S. Alexandrov, Phys. Rev. B **53**, 2863 (1996).
[21] J. Mustre de Leon, I. Batistić, A.R. Bishop, S.D. Conradson, and S.A. Trugman, Phys. Rev. Lett. **68**, 3236 (1992).
[22] H. Fehske, J. Loos, and G. Wellein, Z. Phys. B: Condens. Matter **104**, 619 (1997).
[23] A. La Magna and R. Pucci, Phys. Rev. B **53**, 8449 (1996).
[24] H. Fehske, H. Röder, G. Wellein, and A. Mistriotis, Phys. Rev. B **51**, 16 582 (1995).
[25] P.E. Kornilovitch and E.R. Pike, Phys. Rev. B **55**, R8634 (1997).
[26] M. Capone, W. Stephan, and M. Grilli, Phys. Rev. B **56**, 4484 (1997).
[27] E. Jeckelmann and S.R. White, Phys. Rev. B **57**, 6376 (1998).
[28] E.V. de Mello and J. Ranninger, Phys. Rev. B **58**, 9098 (1998).
[29] J.M. Robin, Phys. Rev. B **58**, 14 335 (1998).
[30] P. Benedetti and R. Zeyher, Phys. Rev. B **58**, 14 320 (1998).
[31] M. Capone, S. Ciuchi, and C. Grimaldi, Europhys. Lett. **42**, 523 (1998).
[32] T. Hakioğlu and M.Y. Zhuravlev, Phys. Rev. B **58**, 3777 (1998).
[33] Y. Zolotaryuk, P.L. Christiansen, and J.J. Rasmussen, Phys. Rev. B **58**, 14 305 (1998).
[34] M. Zoli, Phys. Rev. B **57**, 10 555 (1998).
[35] D.W. Brown, B. West, and K. Lindenberg, Phys. Rev. A **33**, 4110 (1986); V.M. Kenkre and H.-L. Wu, Phys. Rev. B **39**, 6907 (1989); M.I. Salkola, A.R. Bishop, V.M. Kenkre, and S. Raghavan, *ibid.* **52**, R3824 (1995).
[36] A.A. Gogolin, Phys. Status Solidi B **109**, 95 (1982).
[37] A.S. Alexandrov and J. Ranninger, Phys. Rev. B **24**, 1164 (1981); **45**, 13 109 (1992).
[38] W. Stephan, Phys. Rev. B **54**, 8981 (1996).
[39] *Handbook of Mathematical Functions*, edited by M. Abramowitz and I.A. Stegun (Dover, New York, 1965), see p. 230.
[40] M. Zoli, Physica C **317-318**, 528 (1999).
[41] J.M. Robin, Phys. Rev. B **56**, 13 634 (1997).
[42] Yu.A. Firsov, V.V. Kabanov, E.K. Kudinov, and A.S. Alexandrov, Phys. Rev. B **59**, 12 132 (1999).
[43] See *Low Dimensional Cooperative Phenomena*, edited by H.J. Keller (Plenum, New York, 1975).
[44] M.J. Rice, A.R. Bishop, J.A. Krumhansl, and S.E. Trullinger, Phys. Rev. Lett. **36**, 432 (1976); M.J. Rice and N.O. Lipari, *ibid.* **38**, 437 (1977).
[45] H.P.R. Frederikse, W.R. Thurber, and W.R. Hosler, Phys. Rev. **134**, A442 (1964).
[46] A.S. Alexandrov, V.V. Kabanov, and D.K. Ray, Phys. Rev. B **49**, 9915 (1994).

[47] R. Fehrenbacher and M.R. Norman, Phys. Rev. Lett. **74**, 3884 (1995).

[48] T. Timusk, C.C. Homes, and W. Reichardt, in *Anharmonic Properties of High $T_c$ Cuprates*, edited by D. Mihailović *et al.* (World Scientific, Singapore, 1995), p. 171.

[49] S. Lupi, P. Maselli, M. Capizzi, P. Calvani, P. Giura, and P. Roy, Phys. Rev. Lett. **83**, 4852 (1999).

[50] A.H. Romero, D.W. Brown, and K. Lindenberg, Phys. Rev. B **59**, 13 728 (1999).